\documentclass[12pt, onecolumn, letterpaper, journal, draftcls]{IEEEtran}
\usepackage{graphicx, epstopdf}
\usepackage{amssymb}
\usepackage{amsmath}
\usepackage{amsthm}
\usepackage{mathrsfs}
\usepackage{cite}
\usepackage{color}
\usepackage[acronym]{glossaries}
\usepackage[T1]{fontenc}
\usepackage[latin9]{inputenc}
\usepackage{array}
\usepackage{textcomp}
\usepackage{multirow}
\usepackage{soul}
\usepackage{flushend}

\IEEEoverridecommandlockouts

\usepackage{tikz}
\usepackage{textcomp}
\usepackage{hyperref}
\usepackage{lipsum}

\newcommand\copyrighttext{%
  \tiny \textcopyright 2019 IEEE. Personal use of this material is permitted. Permission from IEEE must be obtained for all other uses, in any current or future media, including reprinting/republishing this material for advertising or promotional purposes, creating new collective works, for resale or redistribution to servers or lists, or reuse of any copyrighted component of this work in other works. <DOI: 10.1109/LWC.2019.2894118>}
\newcommand\copyrightnotice{%
\begin{tikzpicture}[remember picture,overlay]
\node[anchor=south,yshift=65pt] at (current page.south) {\fbox{\parbox{\dimexpr\textwidth-\fboxsep-\fboxrule\relax}{\copyrighttext}}};
\end{tikzpicture}%
}
    
\makeatletter
\newcommand*\titleheader[1]{\gdef\@titleheader{#1}}
\AtBeginDocument{%
  \let\st@red@title\@title
  \def\@title{%
    \bgroup\normalfont\small\centering\@titleheader\par\egroup
    \vskip0.5em\st@red@title}
}

\newtheorem{remark}{Remark}
\title{
 Robust Design of AC Computing-Enabled Receiver Architecture for SWIPT Networks
}
\author{
        {
        Ha-Vu Tran, and Georges Kaddoum
        }
\thanks{
Ha-Vu Tran and Georges Kaddoum are with University of
Qu\'{e}bec, \'{E}TS engineering school, LACIME Laboratory, Montreal, Canada  (e-mails: ha-vu.tran.1@ens.etsmtl.ca, georges.kaddoum@etsmtl.ca).
This work has been supported by NSERC discovery grant 435243 - 2013.
}
 }
\begin{document}

\makeatother
\titleheader{\footnotesize{This is the authors'version of the paper that has been accepted for publication in IEEE Wireless Communications Letters. MATLAB codes to reproduce the simulations of this paper are available \href{https://github.com/havutran/Matlab\_WCL/blob/master/Matlab\_SourceCode\_WCL2018-1494.zip}{\textcolor{blue}{here}}.}}
 \maketitle
\copyrightnotice

\begin{abstract}
Inspired by the direct use of alternating current (AC) for computation, we propose a novel integrated information and energy receiver architecture for simultaneous wireless information and power transfer (SWIPT) networks. In this context, the AC computing method, in which wirelessly harvested AC energy is directly used to supply the computing block of receivers, enhances not only computational ability but also energy efficiency over the conventional direct current (DC) one. Further, we aim to manage the trade-off between the information decoding (ID) and energy harvesting (EH) optimally while taking imperfect channel estimation into account. It results in a worst-case optimization problem of maximizing the data rate under the constraints of an EH requirement, the energy needed for supplying the AC computational logic, and a transmit power budget. Then, we propose a method to derive closed-form optimal solutions. The numerical results demonstrate that the proposed architecture with AC computing significantly improves the rate-energy region.

\end{abstract}

\section{Introduction}
Internet of Things (IoT) has been known as an innovative platform for which billions of identified low-power devices, such as sensor nodes, are connected to each other without the need for human interactions \cite{Fuqaha2015}. It is expected that the number of connected IoT devices will exceed 50 billion by 2021 \cite{Sorrell2018}. This gives rise to the challenging issue of replacing the batteries for IoT devices since this task is very time consuming and expensive. Among wireless power transfer approaches using electromagnetic waves \cite{Lu2015}, RF simultaneous wireless information and power transfer (SWIPT) technology, which can support far-field transmission \cite{IoannisKrikidis2014}, have opened up the opportunity for wirelessly recharging and then prolonging the lifetime of IoT networks \cite{IoannisKrikidis2014,Lu2015,Zhang2013,Zhou2013,BrunoClerckx,Tran2017}.

In SWIPT networks, the integrated information and energy receiver architecture of wireless devices, such as power-splitting, and time-switching architectures, plays a critically important role in managing the fundamental trade-off between information decoding (ID) and energy harvesting (EH) performances. Conventionally, the RF signals directed to an energy harvester, having an AC voltage, is rectified to obtain the form of a direct current (DC) voltage. This DC can be used to supply wireless devices. However, the rectified voltage amplitude, determining the computational ability of devices, is often low \cite{Wan2017}. Motivated by this issue, E. Salman {\it et al.} have recently developed a promising approach for computing circuits, that can be directly powered by AC in \cite{Wan2017,Salman2018}.  On this basis, not only the computing efficiency's energy is improved but the part of AC-to-DC conversion loss for computing blocks is also eliminated. The promise of the approach has been demonstrated in previous works \cite{Wan2017,Salman2018}. However, this approach calls for the redesign of existing architectures and managing strategies at the receiver.

In this work, we propose a novel receiver architecture, based on power splitting, which enables AC computing for the SWIPT ID-EH receiver. 
Accordingly, we consider a SWIPT system model where a multi-antenna transmitter conveys information and energy wirelessly to a single antenna receiver using RF signals. 
With the proposed architecture, we aim to derive a strategy to handle the fundamental trade-off between ID and EH performances optimally while accounting for imperfect downlink channel estimation. Hence, it raises an interesting problem on optimizing robust beamformers at the transmitter side and the power splitting ratios for the information decoder, the energy harvester, and the AC computational logic at the receiver side. Also, the data rate is maximized under the constraints of an EH requirement, the energy needed for supplying the AC computational logic, and a transmit power budget. The resulting problem is difficult to solve because it has the form of convex-convex ratio maximization with multi-variable coupling in the constraints. Then, we propose a method to tackle such a problem using closed-form optimal solutions.
The main contributions of this work can be summarized by
\begin{itemize}
\item Proposing the information and energy receiver architecture with AC computational logic for SWIPT networks.
\item Developing closed-form optimal solutions to solve the optimization problem. 
\end{itemize}

\begin{figure}[t]
\centering
{\includegraphics[width=0.55\textwidth]{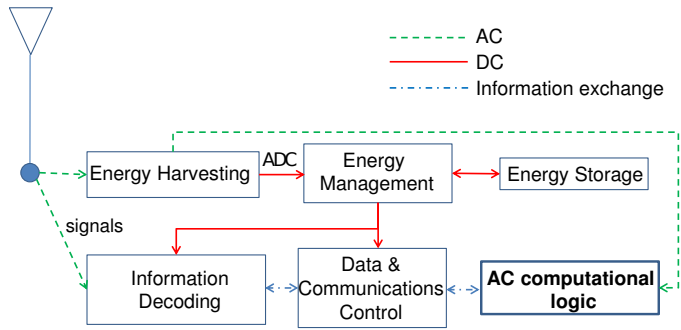}}
   \caption{
       Integrated ID and EH Receiver with AC computing.
    }
    \label{fig:EHModel1}
\end{figure}

\section{System Model}
We consider a SWIPT network where an $M$-antenna transmitter aims to recharge one single-antenna receiver over RF signals.
Particularly, the device can decode information and harvest energy from RF signals. Moreover, it is integrated with a recently proposed AC computational logic \cite{Salman2018}, directly supplied by AC.

More specifically, the high-level architecture of the device can be represented as in Fig. \ref{fig:EHModel1}.
The obtained RF signal is split into two flows for ID and EH. 
Different from conventional energy harvesters \cite{Zhou2013,Lu2015}, the EH module in Fig. \ref{fig:EHModel1} includes not only a voltage multiplier to convert AC to DC but also signal conditioning blocks to produce the required AC signals for AC computational logic \cite{Salman2018}. Further, the energy management block is responsible for handling the task of energy distribution to the ID and the control blocks. Based on the amount of harvested RF energy, the management block decides to draw energy from the storage or to convey excessive energy to the storage for future use.

\begin{figure}[t]
\centering
{\includegraphics[width=0.55\textwidth]{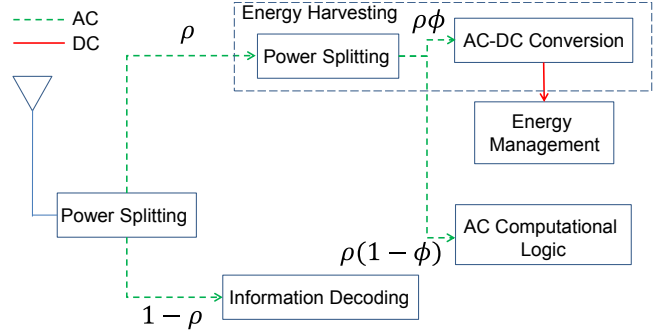}}
   \caption{
      Power splitting-based EH receiver architecture with AC computing.
    }
    \label{fig:EHModel2}
\end{figure}

According to Fig. \ref{fig:EHModel1}, we propose an RF EH receiver architecture with AC computational logic, based on power splitting, as shown in Fig. \ref{fig:EHModel2}. Compared with the conventional power-splitting architecture \cite{Zhou2013,Lu2015}, the proposed one has an extra splitter to direct the AC energy to the AC computational logic.

We assume imperfect channel state information (CSI) is estimated at the transmitter, denoted by $({\mathbf {\hat h}} )^H$ (i.e., ${\mathbf {\hat h}} \in {\mathbb C}^{M}$). Similar to \cite{TuanAnhLe}, the channel model can be modelled as
\begin{align}\label{eq:error}
{\mathbf h} = \mathbf {\hat h} + {\mathbf e},
\end{align}
where $\mathbf {h}$ is the actual channel, ${\mathbf e}$ represents a channel uncertainty defined by
${\mathbf e} \in \mathbb C^{M}$ with $\left\Vert {\mathbf e} \right\Vert ^{2}\le\psi\left\Vert {\mathbf {\hat h}}\right\Vert ^{2}$, where $\psi$ is named as an error factor.
Thus, the received signal is 
\begin{align}\label{eq:RSig}
y &=  ({  {\mathbf h}} )^H   {\mathbf w}  s+ n_0,
\end{align}
where ${\mathbf w}\in {\mathbb C}^{M \times 1}$ is the beamforming vector, $s$ is the complex-valued signal with unit power, modeled as an independent and identically distributed random variable ($s \in { \mathbb C}$) \cite{BrunoClerckx}, and $ n_0$ is the additive white Gaussian noise (AWGN), i.e. $n_0 \sim \mathcal {CN} (0, \sigma^2_{0})$.

According to the architecture given in Fig. \ref{fig:EHModel2}, the data rate (in bps/Hz) can be computed as
\begin{align}\label{eq:rate}
R = \text {log}\left(1 + \dfrac{ \left| ({  {\mathbf h}} )^H   {\mathbf w}\right|^2}{\sigma^2_{0} + \dfrac{\sigma^2_1}{1-\rho}}\right),
\end{align}
where $\sigma^2_{1}$ is the variance of the AWGN noise introduced by the information decoder, and $\rho$ $(0<\rho<1)$ is a power spitting ratio.
Hence, the AC signal power dedicated for supplying the AC computational logic is
\begin{align}\label{eq:AC}
\mathtt {SP}_{AC} =  \rho(1-\phi)\left| ({  {\mathbf h}} )^H   {\mathbf w} \right|^2,
\end{align}
where $\phi$ $(0<\phi<1)$ is a power spitting ratio. Further, we consider the non-linear EH model for DC \cite{Clerckx2016,Boshkovska2017,KeXiong2017,VuTVT2018}, thus the DC EH performance can be calculated as
\begin{align}\label{eq:PCEH2}
\mathtt {EH}_{DC}= \dfrac{ \dfrac{\mathtt{M}^{EH}}{1 + e^{-\mathtt{a}(\mathtt {\hat EH}_{DC} -\mathtt{b})} } - \dfrac{\mathtt{M}^{EH}}{1 + e^\mathtt{ab}} }{1 - \dfrac{1}{1 + e^\mathtt{ab}}},
\end{align}
where 
$\mathtt {\hat EH}_{DC} =  \rho\phi\left| ({  {\mathbf h}} )^H   {\mathbf w} \right|^2$, and $\mathtt{M}^{EH}$ is a constant representing the maximum harvested energy at a user when the RF EH circuit meets saturation. In addition, $\mathtt{a}$ and $\mathtt{b}$ are constants regarding circuit specifications \cite{Clerckx2016,Boshkovska2017,KeXiong2017,VuTVT2018}.

\section{Problem Formulation and Proposed Solution}
\subsection{Worst-case Problem Formulation}
We aim at maximizing the data rate, subject to the constraints of the energy directed to the energy management unit, the energy dedicated for the AC computational logic, and transmit power. Accordingly, the corresponding optimization problem can be formulated as follows:
\begin{subequations}\label{eq:MOProblem1}
\begin{align}
	\text{OP$_1$:} \quad   \underset{{\mathbf w}, \rho, \phi} \max \quad & \underset {\mathbf e} \min \hspace{2pt} R \label{eq:MOProblema} 
\\
\text{s.t.:}  	\quad & \underset {\mathbf e} \min \hspace{2pt} \mathtt {SP}_{AC} \ge \theta \label{eq:MOProblemb}
\\			
			\quad &  \underset {\mathbf e} \min \hspace{2pt} \mathtt {EH}_{DC} \ge \epsilon, \label{eq:MOProblemc}
\\			
			\quad & \left| {\mathbf w} \right|^2 + P_{circ} \le P, \label{eq:MOProblemd}
\\			
			\quad & 0 < \phi < 1, \label{eq:MOProbleme}
\\			
			\quad & 0 < \rho < 1, \label{eq:MOProblemf}						
\end{align}
\end{subequations}
where constraint \eqref{eq:MOProblemb} is to ensure that the energy directed to the AC computational logic is larger than the threshold $\theta$. Also, constraint \eqref{eq:MOProblemc} implies that the energy conveyed to the management unit is larger than the threshold $\epsilon$. Further, $P_{circ}$ is circuit power consumption, and the transmit power is limited by a budget, denoted by $P$, through constraint \eqref{eq:MOProblemd}.

In this context, we denote ${\mathbf w}^{\star}$, $\phi^{\star}$ and $\rho^{\star}$ as the optimal values of ${\mathbf w}$, $\rho$, and $\phi$, respectively. Different from the previous works \cite{Zhang2013,BrunoClerckx}, problem OP$_1$ has one more variable and one more constraint due to taking the AC computing into account, resulting in an intractable form of maximization of the ratio of two convex functions. Furthermore, variables ${\mathbf w}$, $\rho$, and $\phi$ are coupled in constraints \eqref{eq:MOProblemb} and \eqref{eq:MOProblemc}. Hence, solving OP$_1$ is challenging.

\subsection{Proposed Closed-form Optimal Solution}
To solve OP$_1$, we need to have the expressions of $\underset {\mathbf e} \min \hspace{2pt} R$, $\underset {\mathbf e} \min \hspace{2pt} \mathtt {SP}_{AC}$, and  $\underset {\mathbf e} \min \hspace{2pt} \mathtt {EH}_{DC}$ first. Based on \eqref{eq:rate}, \eqref{eq:AC}, and \eqref{eq:PCEH2}, the expressions can be obtained through finding $\underset {\mathbf e} \min \hspace{2pt} \left| ({  {\mathbf h}} )^H   {\mathbf w} \right|^2$, computed below 
\begin{align}\label{eq:WC}
\underset {\mathbf e} \min \hspace{2pt} | ({  {\mathbf h}} )^H   {\mathbf w} |^2 = (1-\sqrt{\psi})^2| ({  {\mathbf {\hat h}}} )^H   {\mathbf w} |^2,
\end{align} 
when ${\mathbf e}=-\sqrt{\psi} \mathbf {\hat h}$. By substituting \eqref{eq:WC} into \eqref{eq:MOProblema}, \eqref{eq:MOProblemb}, and \eqref{eq:MOProblemc}, the expressions of these constraints are then achieved.

Further, to deal with the difficulty of solving problem OP$_1$ mentioned in subsection III.A, we propose a two-step method to deal with it as follows. First, we find the closed-form derivation of the optimal beamformer ${\mathbf w}^{\star}$ while considering $\phi$ and $\rho$ as constants. Second, we substitute the derivation of ${\mathbf w}^{\star}$ into the problem and then optimize $\rho$ and $\phi$. More details are depicted subsequently.

\subsubsection{Finding ${\mathbf w}^{\star}$ for a given $\phi$ and $\rho$}
In the first step, we aim to solve problem OP$_1$ while treating $\phi$ and $\rho$ as constants. For convenience, we reformulate constraints \eqref{eq:MOProblemb} and \eqref{eq:MOProblemc} respectively as
\begin{align}\label{eq:constraintb1}
{| ({  {\mathbf {\hat h}}} )^H   {\mathbf w}|^2} \ge \frac{\theta}{\rho(1-\phi)(1-\sqrt{\psi})^2},
\end{align}
and
\begin{align}\label{eq:constraintc1}
| ({  {\mathbf {\hat h}}} )^H   {\mathbf w} |^2 \ge \frac{\bar \epsilon}{\rho\phi(1-\sqrt{\psi})^2},
\end{align}
where
$\bar \epsilon = \mathtt{b} - \frac{1}{\mathtt{ a}}\text{ln}\left(  \frac{e^{\mathtt {ab}}({\mathtt M}^{EH} - \epsilon)}{e^\mathtt{ ab}\epsilon + \mathtt M^{EH} } \right).$

Since log is an increasing function, problem OP$_1$ can be equivalently represented as
\begin{align}
	\text{SubOP$_1$:} \quad &  \underset{{\bf w}} \max \quad  | ({  {\mathbf {\hat h}}} )^H   {\mathbf w} |^2  \nonumber
\\
\text{s.t.:}  	\quad & \eqref{eq:constraintb1}, \hspace{3pt} \eqref{eq:constraintc1}, \hspace{3pt} \eqref{eq:MOProblemd}, \hspace{3pt} \text{and} \hspace{3pt} \eqref{eq:MOProbleme}.
\end{align}
For given constants $\phi$, $\rho$, $\theta$, and $\bar \epsilon$, it can be seen that SubOP$_1$ is feasible if and only if there is a large enough value of $P$ which allows $| ({  {\mathbf {\hat h}}} )^H   {\mathbf w} |^2$ to satisfy constraints \eqref{eq:constraintb1} and \eqref{eq:constraintc1}.

We continue solving SubOP$_1$ by assuming that SubOP$_1$ is feasible. It implies that constraints \eqref{eq:constraintb1} and \eqref{eq:constraintc1} are satisfied and then can be neglected. Therefore, SubOP$_1$ simply becomes the maximization of $| ({  {\mathbf {\hat h}}} )^H   {\mathbf w} |^2$ with $| {\mathbf w} |^2 = P - P_{circ}$. Thus, ${\mathbf w}^{\star}$ can be computed by the following closed-form expression
\begin{align}
{\mathbf w}^{\star} = \sqrt{P - P_{circ}}  {\mathbf v},
\end{align}
where ${\mathbf v}$ is the orthonormal eigenvector corresponding to the largest eigenvalue $\lambda_\text{largest} \left( {  {\mathbf {\hat h}}} ({  {\mathbf {\hat h}}})^H \right)$. Thus, we can draw a significant remark as follows
\begin{remark}
It is found that the optimal beamformer ${\mathbf w}^{\star}$ can be derived as a closed-form expression not including $\rho$ and $\phi$. In other words, we could have ${\mathbf w}^{\star}$ without solving $\rho^{\star}$ and $\phi^{\star}$. 
\end{remark}

\subsubsection{Finding $\rho^{\star}$ and $\phi^{\star}$ with solved ${\mathbf w}^{\star}$}
In the second step, by substituting ${\mathbf w}^{\star}$ into problem OP$_1$, we then find $\rho^{\star}$ and $\phi^{\star}$.
Initially, for convenience, let
$(1-\sqrt{\psi})^2\left| ({  {\mathbf {\hat h}}} )^H   {\mathbf w}^{\star} \right|^2 = (1-\sqrt{\psi})^2({P - P_{circ}}) \left| ({  {\mathbf {\hat h}}} )^H  {\mathbf v} \right|^2 = \Gamma,$ where $\Gamma $ is a constant. Next, we reformulate constraints \eqref{eq:MOProblemb} and \eqref{eq:MOProblemc}, respectively, as
\begin{align}\label{eq:reformb}
\rho \ge  \dfrac{\theta}{\Gamma(1-\phi)},
\end{align}
and
\begin{align}\label{eq:reformc}
\rho &\ge \dfrac{\bar \epsilon}{\phi \Gamma}.
\end{align}

By combining \eqref{eq:reformb} and \eqref{eq:reformc}, we have
\begin{align}\label{eq:reformd}
\rho \ge \max \left\{ \tfrac{\theta}{\Gamma(1-\phi)}, \tfrac{\bar \epsilon}{\phi \Gamma}  \right\}.
\end{align}

Considering objective \eqref{eq:MOProblema}, the objective can be seen as maximizing $1-\rho$. Furthermore, maximizing $1-\rho$ with $0 < \rho < 1$ is equivalent to minimizing $\rho$. Accordingly, also taking \eqref{eq:reformd} into account, problem OP$_1$ can be equivalently transformed into OP$_2$ as
\begin{align}
	\text{OP$_2$:} \quad &  \underset{\phi} \min \quad \underset{ \phi}\max \left\{ \tfrac{\theta}{\Gamma(1-\phi)}, \tfrac{\bar \epsilon}{\phi \Gamma}  \right\} \nonumber
\\
\text{s.t.:}  	\quad & \eqref{eq:MOProbleme}.
\end{align}
In general, problem OP$_2$ is still in an intractable formulation to solve. However, since $0 < \phi < 1$ and the objective function is in the form of $\max \left\{ f( \frac{1}{1-\phi}), g(\frac{1}{\phi}) \right\}$, we have found an interesting property for problem OP$_2$, presented as follows.
\newtheorem{lem}{Lemma}
\begin{lem}
The optimal value of $\phi$, denoted by $\phi^{\star}$, should satisfy the equation below
\begin{align}
 \dfrac{\theta}{(1-\phi^{\star})  \Gamma} = \dfrac{\bar \epsilon}{\Gamma\phi^{\star}}.
\end{align}
\end{lem}
\begin{IEEEproof}
See Appendix A.
\end{IEEEproof}

Thus, according to {\it Lemma} 1, the closed-form optimal solutions of OP$_2$ can be calculated as
\begin{align}
\phi^{\star} = \dfrac{\bar \epsilon}{\theta + \bar \epsilon} ,
\end{align}
and then
\begin{align}
\rho^{\star} = \dfrac{\bar \epsilon}{\Gamma\phi^{\star}}.
\end{align}
To this end, we provide an important remark below
\begin{remark}
It can be seen that every step solving OP$_1$ is derived without the loss of generality. Then, our method does not yield any optimality loss. Notably, it is clear that our closed-form solutions only require simple calculations.

For the extension of multiple users, since the closed-form solution of beamformers may not be achieved, it is suggested that an iterative algorithm may be required to update one variable while fixing the others until convergence.
\end{remark}

\section{Numerical Results}
In this section, the number of transmit antenna is set to $M=4$, and the receiver is located $4$m far away from the transmitter. Specifically, the channels between the transmitter and the user are assumed to have Rician distribution in which the Rician factor is set to 6 dB and the pathloss exponent factor is set to 2.6. Moreover, we set $\sigma_0^2 = -111$ dBm and $\sigma_1^2 = 35$ dBm. Regarding the nonlinear EH model, we set $\mathtt{M}^{EH} = 3.9$ mW, $\mathtt{a} = 1500$ and $\mathtt{b} = 0.0022$ \cite{Boshkovska2017,KeXiong2017}. Particularly, we set $\theta = 0.04764$ mW and $\theta =  0.00027$ mW \cite{Salman2018} which are the energy needed for supplying conventional DC computing and AC computing blocks, respectively. The simulation is carried out using 10000 channel realizations.

\begin{figure}[t]
\centering
{\includegraphics[width=0.6\textwidth]{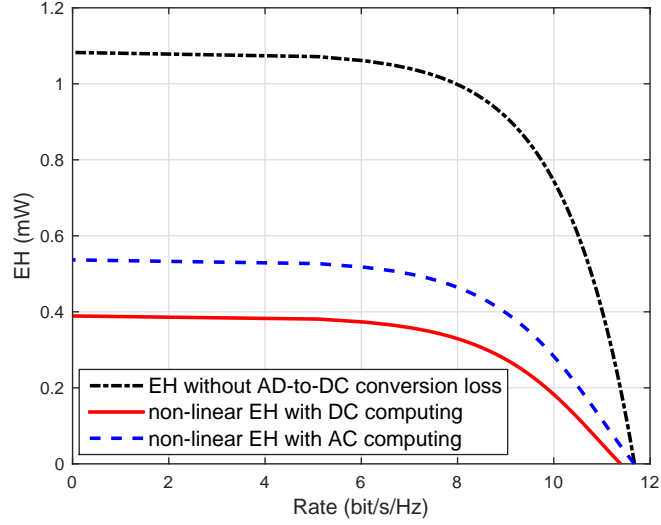}}
   \caption{
       Rate-Energy region, $P_{0} = P - P_{circ} = 10$ dBm, $\psi = 0$.
    }
    \label{fig:sim1}
\end{figure}

In Fig. \ref{fig:sim1}, a performance comparison between the conventional architecture and the one with AC computing is shown under the perfect CSI condition, for a fair observation. It is observed that using the proposed AC computing-enabled architecture results in a significantly larger rate-energy region. It can be explained by the facts that (i) there is no AC-DC conversion needed for the AC computational logic, and (ii) supplying the AC computing block also requires less energy than the DC computing one. This outcome is consistent with the measurement results in \cite{Wan2017,Salman2018} where the AC computing outperforms the DC one.

\begin{figure}[t]
\centering
{\includegraphics[width=0.6\textwidth]{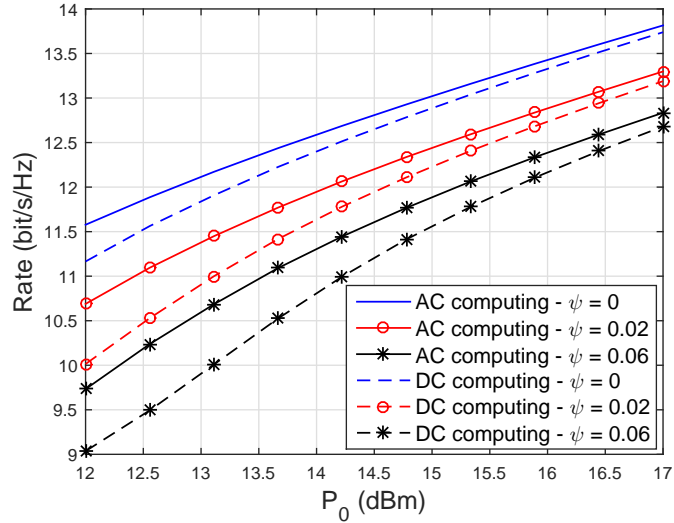}}
   \caption{
       Impact of imperfect CSI, $P_{0} = P - P_{circ}$ and $\epsilon = 0.2$.
    }
    \label{fig:sim2}
\end{figure}

Fig. \ref{fig:sim2} presents the impact of channel error factor $\psi$ on the data rate in the cases with AC and DC computing blocks. It can be seen that the data rate increases as the transmit power scales up. 
Further, at the same transmit power, addressing a higher $\psi$ results in a lower data rate. It is because all the constraints require more energy to be satisfied, and then less energy is dedicated to the ID purpose.

\section{Conclusion}
In this work, we proposed a novel integrated information and energy receiver architecture, based on power splitting, to enable the AC computing methodology for SWIPT users. Accounting for the imperfect CSI, we devised the robust strategy to maximize the data rate under the constraints of the harvested AC and DC energies and the transmit power budget. We provided a simple closed-form optimal solution to the problem to facilitate the solving process. The numerical results indicate that the proposed architecture significantly improves the rate-energy region. Therefore, it can be evaluated as one of the principal candidates in developing smarter IoT devices with more powerful computational capability and longer lifetime.

\appendices
\section{Proof of Lemma 1}

We assume that there is a value of $\phi_0$ such that
\begin{align}
 \dfrac{\theta}{(1-\phi_0)\Gamma}  = \dfrac{\bar \epsilon}{\Gamma\phi_0} = \Xi.
\end{align}

Next, by increasing $\phi_0$ with $a$ ($ a > 0$ and $0 < a + \phi_0 < 1$), it can be observed that
\begin{align}\label{eq:lemma11}
&\max \left\{  \dfrac{\theta}{\Gamma (1-(\phi_0+a)) } , \dfrac{\bar \epsilon}{\Gamma(\phi_0+a)}  \right\} = \dfrac{\theta}{\Gamma (1-(\phi_0+a)) }  > \Xi.
\end{align}

Furthermore, by decreasing $\phi_0$ with $b$ ($ b > 0$ and $0 < \phi_0 - b < 1$)
\begin{align}\label{eq:lemma12}
&\max \left\{  \frac{\theta}{ \Gamma (1-(\phi_0-b))} , \frac{\bar \epsilon}{\Gamma(\phi_0-b)}  \right\} = \frac{\bar \epsilon}{\Gamma(\phi_0-b)} > \Xi.
\end{align}

Thus, \eqref{eq:lemma11} and \eqref{eq:lemma12} imply that either increasing $\phi$ or decreasing $\phi$ results in a scale up of the objective function of OP$_3$. Therefore, $\phi_0$ is the optimal solution of OP$_3$ (i.e., $\phi^{\star} = \phi_0$), and the proof of {\it Lemma} 1 is completed.

\bibliographystyle{IEEEtran}
\bibliography{IEEEabrv,Tran_WCL2018-1494_REF}

\begin{thebibliography}{10}
\providecommand{\url}[1]{#1}
\csname url@samestyle\endcsname
\providecommand{\newblock}{\relax}
\providecommand{\bibinfo}[2]{#2}
\providecommand{\BIBentrySTDinterwordspacing}{\spaceskip=0pt\relax}
\providecommand{\BIBentryALTinterwordstretchfactor}{4}
\providecommand{\BIBentryALTinterwordspacing}{\spaceskip=\fontdimen2\font plus
\BIBentryALTinterwordstretchfactor\fontdimen3\font minus
  \fontdimen4\font\relax}
\providecommand{\BIBforeignlanguage}[2]{{%
\expandafter\ifx\csname l@#1\endcsname\relax
\typeout{** WARNING: IEEEtran.bst: No hyphenation pattern has been}%
\typeout{** loaded for the language `#1'. Using the pattern for}%
\typeout{** the default language instead.}%
\else
\language=\csname l@#1\endcsname
\fi
#2}}
\providecommand{\BIBdecl}{\relax}
\BIBdecl

\bibitem{Fuqaha2015}
A.~Al-Fuqaha, M.~Guizani, M.~Mohammadi, M.~Aledhari, and M.~Ayyash, ``Internet
  of things: A survey on enabling technologies, protocols, and applications,''
  \emph{IEEE Comm. Surveys \& Tutorials}, vol.~17, no.~4, pp. 2347 -- 2376,
  Fourthquarter 2015.

\bibitem{Sorrell2018}
\BIBentryALTinterwordspacing
S.~Sorrell, \emph{The Internet of Things: Consumer, Industrial \& Public
  Services 2018-2023}.\hskip 1em plus 0.5em minus 0.4em\relax Juniper, 2018.
  [Online]. Available:
  \url{https://www.juniperresearch.com/press/press-releases/iot-connections-to%
-grow-140-to-hit-50-billion}
\BIBentrySTDinterwordspacing

\bibitem{Lu2015}
X.~Lu, P.~Wang, D.~Niyato, D.~I. Kim, and Z.~Han, ``Wireless networks with {RF}
  energy harvesting: A contemporary survey,'' \emph{IEEE Comm. Surveys \&
  Tutorials}, vol.~17, no.~2, pp. 757--789, Secondquarter 2015.

\bibitem{IoannisKrikidis2014}
I.~Krikidis, S.~Timotheou, S.~Nikolaou, G.~Zheng, D.~W.~K. Ng, and R.~Schober,
  ``Simultaneous wireless information and power transfer in modern
  communication systems,'' \emph{{IEEE} Commun. Mag.}, vol.~52, no.~11, pp. 104
  --110, Nov. 2014.

\bibitem{Zhang2013}
R.~Zhang and C.~K. Ho, ``{MIMO} broadcasting for simultaneous wireless
  information and power transfer,'' \emph{{IEEE} Trans. Wireless Commun.},
  vol.~12, no.~5, pp. 1989--2001, 2013.

\bibitem{Zhou2013}
X.~Zhou, R.~Zhang, and C.~K. Ho, ``Wireless information and power transfer:
  Architecture design and rate-energy tradeoff,'' \emph{IEEE Transactions on
  Communications}, vol.~61, no.~11, pp. 4754--4767, 2013.

\bibitem{BrunoClerckx}
B.~Clerckx, R.~Zhang, R.~Schober, D.~W.~K. Ng, D.~I. Kim, and H.~V. Poor,
  ``Fundamentals of wireless information and power transfer: From {RF} energy
  harvester models to signal and system designs,'' \emph{{IEEE} J. Sel. Areas
  Commun.}, vol.~37, no.~1, pp. 4--33, Jan. 2019.

\bibitem{Tran2017}
H.-V. Tran and G.~Kaddoum, ``{RF} wireless power transfer: Regreening future
  networks,'' \emph{IEEE Potentials}, vol.~37, no.~2, pp. 35 -- 41, Mar.-Apr.
  2018.

\bibitem{Wan2017}
T.~Wan, Y.~Karimi, M.~Stanacevic, and E.~Salman, ``Perspective paper - can {AC}
  computing be an alternative for wirelessly powered {IoT} devices?''
  \emph{IEEE Embed. Syst. Lett.}, vol.~9, no.~1, pp. 13 -- 16, Mar. 2017.

\bibitem{Salman2018}
E.~Salman, M.~Stanacevic, S.~Das, and P.~M. Djuric, ``Leveraging {RF} power for
  intelligent tag networks,'' in \emph{ACM Great Lakes Symposium on VLSI},
  Chicago, IL, USA, May 2018, pp. 329--334.

\bibitem{TuanAnhLe}
T.~A. Le, Q.-T. Vien, H.~X. Nguyen, D.~W.~K. Ng, and R.~Schober, ``Robust
  chance-constrained optimization for power-efficient and secure {SWIPT}
  systems,'' \emph{IEEE Transactions on Green Communications and Networking},
  vol.~1, no.~3, pp. 333 -- 346, May 2017.

\bibitem{Clerckx2016}
B.~Clerckx and E.~Bayguzina, ``Waveform design for wireless power transfer,''
  \emph{{IEEE} Trans. Signal Process.}, vol.~64, no.~23, pp. 6313--6328, Dec.
  2016.

\bibitem{Boshkovska2017}
E.~Boshkovska, D.~W.~K. Ng, N.~Zlatanov, A.~Koelpin, and R.~Schober, ``Robust
  resource allocation for {MIMO} wireless powered communication networks based
  on a non-linear {EH} model,'' \emph{{IEEE} Trans. Commun.}, vol.~65, no.~5,
  pp. 1984 -- 1999, May 2017.

\bibitem{KeXiong2017}
K.~Xiong, B.~Wang, and K.~J.~R. Liu, ``Rate-energy region of {SWIPT} for {MIMO}
  broadcasting under nonlinear energy harvesting model,'' \emph{{IEEE} Trans.
  Wireless Commun.}, vol.~16, no.~8, pp. 5147 -- 5161, May 2017.

\bibitem{VuTVT2018}
H.-V. Tran, G.~Kaddoum, and K.~T. Truong, ``Resource allocation in {SWIPT}
  networks under a nonlinear energy harvesting model: Power efficiency, user
  fairness, and channel nonreciprocity,'' \emph{{IEEE} Trans. Veh. Technol.},
  vol.~67, no.~9, pp. 8466 -- 8480, Sept. 2018.

\end{thebibliography}
\end{document}